\documentclass[a4paper]{jpconf}
\usepackage{graphicx}
\usepackage{amsmath,amssymb}

\begin{document}
\title{The Initial Flow of Classical Gluon Fields in Heavy Ion Collisions}

\author{Rainer J Fries and Guangyao Chen}

\address{Cyclotron Institute and Department of Physics and Astronomy, Texas
  A\&M University, College Station TX 77843, USA}

\ead{rjfries@comp.tamu.edu}

\begin{abstract}
Using analytic solutions of the Yang-Mills equations we calculate the 
initial flow of energy of the classical gluon field created in collisions 
of large nuclei at high energies. We find radial and elliptic flow which
follows gradients in the initial energy density, similar to a simple
hydrodynamic behavior. In addition we find a rapidity-odd transverse flow
field which implies the presence of angular momentum and should lead to
directed flow in final particle spectra. We trace those energy flow 
terms to transverse fields from the non-abelian generalization of Gauss' Law
and Amp\'ere's and Faraday's Laws.
\end{abstract}

\section{Introduction}

At asymptotically large energies (or small Bjorken-$x$) the gluon fields 
in hadrons and nuclei approach a state of nuclear matter generally referred to
as color glass condensate (CGC) \cite{Iancu:2003xm,Gelis:2010nm}. The 
transverse gluon density saturates at a value $\sim Q_s^{-2}$ 
characterized by a saturation scale $Q_s$. In this limit large occupation 
numbers permit a quasi-classical treatment of the gluon field which 
is the approximation known as the McLerran-Venugopalan (MV) model 
\cite{McLerran:1993ka,McLerran:1993ni}. The saturation scale $Q_s$ grows with 
the size of the nucleus $\sim A^{1/3}$. Hence high energy nuclear 
collisions at the Relativistic Heavy Ion Collider (RHIC) and the 
Large Hadron Collider (LHC) offer unique opportunities to study CGC.

Here we report on results from a calculation which solves the classical
gluon field (Yang-Mills) equations employing an expansion in powers of the
proper time $\tau = \sqrt{t^2-z^2}$ after the collision. We focus on the 
transverse Poynting vector $S^i = T^{0i}$, $i=1,2$ of the gluon field, 
where $T^{\mu\nu}$ is the energy momentum tensor. $S^i$ describes the initial 
transverse flow of energy of the gluon field, but we expect this flow to 
translate into hydrodynamic flow of quark gluon plasma once the system
thermalizes. Our results have first been reported in detail in 
\cite{Chen:2012es,Chen:2013ksa}.

\section{Color Glass Condensate}

We seek solutions of the classical Yang-Mills equations
\begin{equation}
  \left[ D_\mu, F^{\mu\nu}\right] = J^\nu
  \label{eq:ym}
\end{equation}
where the $SU(3)$-current $J^\mu$ is generated by two transverse 
charge distributions $\rho_1(\vec x_\perp)$ and $\rho_2(\vec x_\perp)$ 
moving on the $+$ and $-$ light cone respectively. 
The $\rho_i$ describe the $SU(3)$ charge distributions of 
the large-$x$ partons in the colliding nuclei which generate the small-$x$
gluon field. Typically these equations are solved numerically in the 
forward light cone ($\tau \ge 0$) 
\cite{Krasnitz:2000gz,Lappi:2003bi,Schenke:2012wb}. 

However, it is possible to obtain an analytic recursive solution 
as well, as first described in \cite{Fries:2006pv}.
To that end one employs a power series for the gauge potential $A^\mu$
in the forward light cone,
\begin{align}
  A(\tau,\vec x_\perp) &= \sum_{n=0}^\infty \tau^n
  A_{(n)}(\vec x_\perp) \, , \\
  A_\perp^i(\tau,\vec x_\perp) &= \sum_{n=0}^\infty \tau^n
  A_{\perp(n)}^i (\vec x_\perp) \, ,
\end{align}
where in light cone notation $A^\pm = \pm x^\pm A$. The leading terms
are given by boundary conditions on the light cone \cite{KoMLeWei:95} 
\begin{align}
  A_{\perp(0)}^i  &= A_1^i  + A_2^i \, ,
  \label{eq:bc_boost4}\\
  A_{(0)} &= -\frac{ig}{2} \left[ A_1^i , A_2^i \right] \, ,
  \label{eq:bc_boost5}
\end{align}
where $A^i_1$ and $A_2^i$ are the gauge potentials in nucleus 1 and 2
(generated by charges $\rho_1$ and $\rho_2$), respectively, 
before the collision.
The recursion relation for coefficients of even powers $n$ ($n>1$), are
\begin{align}
  A_{(n)} =& \frac{1}{n(n+2)} \sum_{k+l+m=n-2} \left[ D^i_{(k)}, \left[
  D^i_{(l)}, A_{(m)} \right] \right] ,  \label{eq:a_recursion1} \\
  A^i_{\perp(n)} =& \frac{1}{n^2}\left( \sum_{k+l=n-2}
  \left[ D^j_{(k)}, F^{ji}_{(l)} \right]   +
  ig \sum_{k+l+m=n-4} \left[ A_{(k)}, [ D^i_{(l)},A_{(m)} ] \right] \right)
  \, ,  \label{eq:a_recursion2}
\end{align}
while all odd coefficients vanish. The convergence radius of this series
in $\tau$ will generally be of order $1/Q_s$, but for example the series
recovers the known solution for the abelian case for all times \cite{Fries:new}.

The physical fields $F^{\mu\nu}$ can be expanded in $\tau$ as well and can
be computed order by order from $A^\mu$. The dominant fields for small 
times, i.e.\ at order $\mathcal{O}(\tau^0)$, are the longitudinal 
chromo-electric and -magnetic fields \cite{Fries:2006pv,Lappi:2006fp}
\begin{align}
  \label{eq:e0}
  E_{(0)} = F^{+-}_{(0)} &= ig \left[ A_1^i,
  A_2^i \right] \, ,\\
  \label{eq:b0}
  B_{(0)} = F^{21}_{(0)} &= ig \epsilon^{ij} \left[ A_1^i,
  A_2^j \right] \, .
\end{align}
From this result one can calculate an initial energy density 
$\epsilon_0 = T^{00}_{(0)}$. After averaging over charge densities $\rho_i$
one obtains an event-averaged energy density \cite{Lappi:06,Fries:new}
\begin{equation}
 \varepsilon_0 (\vec x_\perp)=    
 2\pi \alpha_s^3\frac{N_c}{ N_c^2-1 } \mu_1 (\vec x_\perp) \mu_2 (\vec x_\perp) \ln^2 \frac{Q^2}{\hat m^2} \, .
\end{equation}
Here the usual assumptions of the MV model have been applied, i.e.\ 
the $\rho_i$ follow Gaussian distributions and $\mu_1$ and $\mu_2$
determine the average squared charge distribution in each nucleus.
$Q$ and $\hat m$ are UV and IR cutoffs respectively.

\section{Transverse Fields and Transverse Flow}

\begin{figure}[h]
\includegraphics[width=16pc]{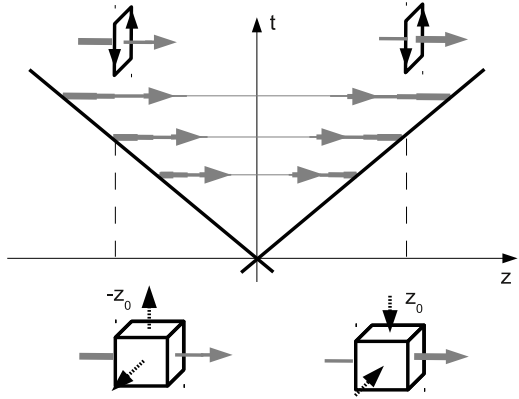}\hspace{2pc}%
\begin{minipage}[b]{16pc}\caption{\label{gaussampere}Schematic sketch of 
Gauss' Law (cubes) and Amp\'ere's and Faraday's Laws (loops)
for positive and negative values of the longitudinal coordinate $z$. 
The initial longitudinal electric and magnetic fields in the forward light 
cone diminish as functions of $t$ and $z$. Transverse fields due to 
Gauss' Law will have different signs for positive and negative $z$, while 
transverse fields due to Amp\'ere's and Faraday's Law are symmetric in 
$z$. The abelian version is shown here, see \cite{Chen:2013ksa} for details.}
\end{minipage}
\end{figure}

Transverse electric and magnetic fields enter at order 
$\mathcal{O}(\tau^1)$ in the power series of $F^{\mu\nu}$.
The corresponding coefficients can be computed to be \cite{Chen:2013ksa}
\begin{align}
  E^i_{(1)} &= -\frac{1}{2} \left( \sinh\eta [D^i, E_0] + \cosh\eta \,
  \epsilon^{ij}[D^j,B_0] \right)
  \label{eq:trfield1}  \\
  B^i_{(1)} &= \frac{1}{2} \left( \cosh\eta\, \epsilon^{ij} [D^j, E_0]
  - \sinh\eta [D^i,B_0] \right) \, .
  \label{eq:trfield2}
\end{align}
Here $\eta$ is the space-time rapidity.
One can readily verify that these expressions are simply the $SU(3)$ analogs
of Gauss' Law and Faraday's and Amp\'ere's Laws. More specifically,
the $\eta$-odd terms emerge naturally as a consequence of Gauss' Law, see
Fig.\ \ref{gaussampere}.

\begin{figure}[h]
\includegraphics[width=16pc]{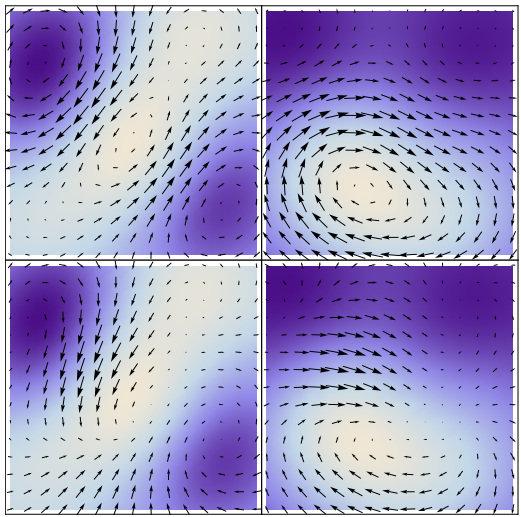}\hspace{2pc}
\begin{minipage}[b]{16pc}
\caption{\label{randomfield} Transverse electric fields (left panels) 
and magnetic fields (right panels) at $\eta = 0$ (upper panels) and 
$\eta = 1$ (lower panels) for a random distribution of initial longitudinal 
electric (shown as background in the right panels) and magnetic fields (left 
panels). 
At $\eta = 0$ the fields are divergence-free. The abelian case is shown here,
see \cite{Chen:2013ksa} for details.}
\end{minipage}
\end{figure}

Fig.\ \ref{randomfield} shows a typical configuration of transverse fields 
(in the abelian case for simplicity) for randomly seeded longitudinal fields
$E_0$ and $B_0$. At rapidity $\eta=0$ the transverse fields are
divergence-free, i.e.\ only closed field lines due to Amp\'ere's and 
Faraday's Law appear. At $\eta=1$ contributions from Gauss' Law are present 
as well.

We are now able to calculate the flow of energy due to transverse fields.
The transverse Poynting vector $S^i$ receives its first contribution in the
power series of $T^{\mu\nu}$ at order $\mathcal{O}(\tau^1)$. I.e.\ 
transverse flow in color glass sets in linearly in time, 
\begin{equation} 
 \label{eq:T0i_1}
  T^{0i} = \frac{\tau}{2} \cosh\eta \, \alpha^i +
     \frac{\tau}{2} \sinh\eta \, \beta^i \, .
\end{equation}
There is a rapidity-even part $\alpha^i$ and a rapidity-odd contribution 
$\beta^i$ to the Poynting vector which are given by \cite{Chen:2013ksa}
\begin{equation}
  \alpha^i = - \nabla^i \varepsilon_0 \quad \, , \qquad
  \beta^i = \epsilon^{ij} \left( [D^j,B_0]E_0 - [D^j,E_0]B_0\right) \, .
\end{equation}
The rapidity-odd flow is expected from the existence of rapidity-odd gauge 
fields, however it is more of a surprise when one approaches the topic of
early flow from a purely phenomenological point of view. $\alpha^i$ corresponds
to energy flow as expected from hydrodynamic expansion, following the
gradient of energy. On the other hand $\beta^i$ is determined by the gauge
field structure underlying the energy density. 

\begin{figure}[h]
\includegraphics[width=20pc]{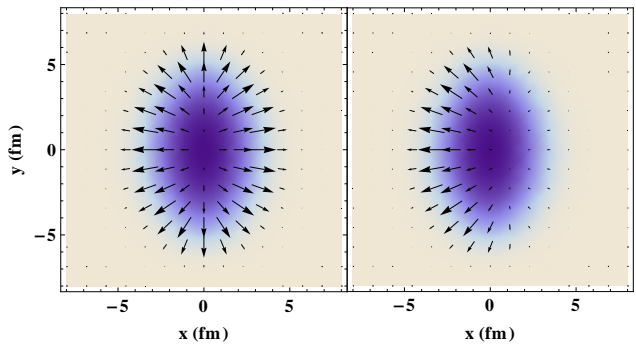}\hspace{2pc}
\begin{minipage}[b]{16pc}
\caption{\label{auautrans} Transverse flow of energy $T^{0i}$ (black arrows;
arbitrary units) and energy density $\varepsilon_0$ (shading) in the 
transverse plane for 
Au+Au collisions at impact parameter $b = 6$ fm. The nucleus with center
located at $x=3$ fm travels into the plane (the positive $\eta$-direction). 
Left Panel: $\eta =0$. Right Panel: $\eta = 1$.
}
\end{minipage}
\end{figure}

After averaging over color charges $\rho_i$ we obtain event-averaged
transverse flow fields \cite{Chen:2013ksa}
\begin{equation}
  \alpha^i =  
  -\varepsilon_0  \frac{\nabla^i\left( \mu_1 \mu_2 \right)}{\mu_1 \mu_2}
  \quad , \qquad   
  \beta^i = - \varepsilon_0 \frac{\mu_2 \nabla^i \mu_1 - \mu_1 \nabla^i
    \mu_2}{\mu_1\mu_2}   \, .
\end{equation}
We omit the notation $\langle \ldots \rangle$ for event-averaged quantities
if no confusion can arise. Fig.\ \ref{auautrans} shows the initial 
event-averaged transverse flow field in Au+Au collisions at impact parameter 
$b=6$ fm for two space-time rapidities $\eta$.

\begin{figure}[h]
\includegraphics[width=12pc]{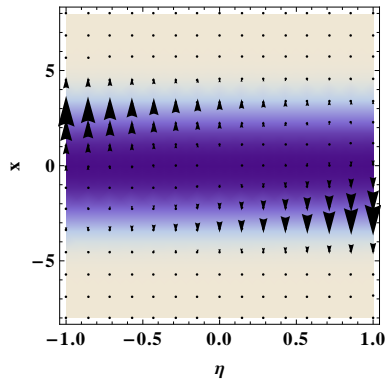}\hspace{2pc}
\begin{minipage}[b]{16pc}
\caption{\label{auauy0} Same as Fig.\ \ref{auautrans} but plotted in the 
$\eta - x$-plane defined by $y = 0$. A flow pattern akin to directed flow
emerges. The nucleus with center at $x=3$ fm is traveling to the right.
}  
\end{minipage}
\end{figure}

One can clearly see a radial and elliptic flow pattern for 
$\eta=0$ which is qualitatively similar to what would develop in 
hydrodynamics. At $\eta=1$ the rapidity-odd flow $\beta^i$ becomes 
comparable to $\alpha^i$ and we notice that the gluon energy 
flow develops a preferred direction. This is akin
to directed flow $v_1$. We can see this more clearly in Fig.\ \ref{auauy0}
where the transverse flow is shown as a function of $x$ and $\eta$ in the
plane $y=0$. The gluon field expands more rapidly in the wake of a passing
nucleus.

\begin{figure}[h]
\includegraphics[width=20pc]{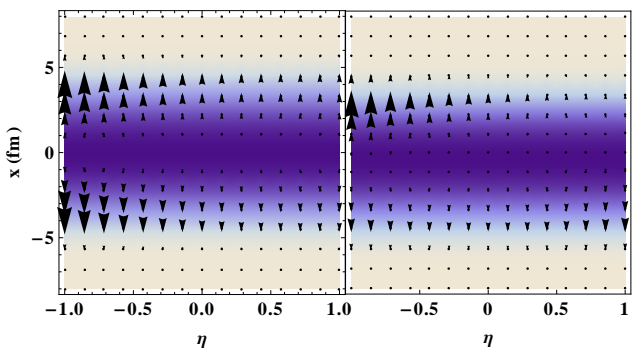}\hspace{2pc}
\begin{minipage}[b]{14pc}
\caption{\label{aucu}The same as Fig.\ \ref{auauy0} for Au+Cu (Au traveling to
the right). Left Panel: $b = 0$ fm. Right Panel: $b = 2$ fm.}  
\end{minipage}
\end{figure}

Interesting flow patterns can also be observed for asymmetric collision 
systems. Fig.\ 
\ref{aucu} shows the transverse flow in the $x-\eta$-plane for Au+Cu
collisions for two different impact parameters. Once again energy flow
is larger in the wake of the spectators of the larger nucleus, 
leading to a strong asymmetry between forward and backward rapidities.

\section{Summary and Discussion}

We have calculated the initial flow of energy of the classical gluon field
in high energy nuclear collisions. Our results are valid up to roughly 
a time $\sim 1/Q_s$ after the collision. The 
gluon field is expected to decohere and thermalize soon after. Energy
and momentum conservation will translate the flow fields found here into
hydrodynamic flow which then will develop further and eventually freeze out 
into particle flow. We expect the rapidity-odd pre-equilibrium flow $\beta^i$
to translate into (rapidity-odd) directed flow $v_1$ of particles. The
sign and general rapidity dependence of $\beta^i$ are qualitatively 
consistent with experimental results \cite{Chen:2013ksa} but a more 
quantitative statement would require a follow-up 3+1-D viscous 
hydrodynamic simulation.

One can also interpret the flow term $\beta^i$ for symmetric collision systems 
with finite impact parameters $b$ with the inevitable presence of 
angular momentum $L_y$ (perpendicular to the reaction plane).
The initial gluon field transfers a part of the angular momentum in
the system before the collision onto the fireball after the collision
\cite{Liang:2004ph,Csernai:2011gg}.  There it leads to a rotation of the 
fireball and possibly to vorticity in the quark gluon fluid.

Another very interesting result is the flow found for asymmetric A+B 
collision systems. The field $\beta^i$ specifically relies on the fact that 
classical gauge fields are the relevant degrees of freedom and one could 
speculate that observables exist in asymmetric collision systems which are 
unique signatures for the flow of gauge fields. A further investigation into 
this direction would be worthwhile.

In the future we plan to use our results on pre-equilibrium flow in
viscous hydrodynamic calculations. Preliminary results show that key features
of $T^{\mu\nu}$, like angular momentum, readily translate into hydrodynamic
fields (local energy density,fluid velocity, shear stress, etc.) in a 
rapid thermalization scenario \cite{Fries:2005yc,Schenke:2012wb}.

\ack
This work was supported by the U.S. National Science Foundation through 
CAREER grant PHY-0847538, and by the JET Collaboration and DOE grant
DE-FG02-10ER41682.

\end{document}